\documentclass[prl,twocolumn,superscriptaddress]{revtex4-2}
\usepackage[dvipdfmx]{graphicx} 
\usepackage{hyperref}
\usepackage{dcolumn}
\usepackage{bm}
\usepackage{color}
\usepackage{ulem}
\usepackage{lineno}

\graphicspath{{fig/}}

\begin{document}
\preprint{APS/123-QED}

\title{
Theoretical study on the possibility of high $T_c$ $s\pm$-wave superconductivity in the heavily hole-doped infinite layer nickelates 
} 
\author{Hirofumi Sakakibara}
\email{sakakibara@tottori-u.ac.jp}
\affiliation{Advanced Mechanical and Electronic System Research Center(AMES), Faculty of Engineering, Tottori University, 4-10 Koyama-cho, Tottori, Tottori 680-8552, Japan}
\author{Ryota Mizuno}
\affiliation{Forefront Research Center, University of Osaka, 1-1 Machikaneyama-cho, Toyonaka, Osaka 560-0043, Japan}
\author{Masayuki Ochi}
\affiliation{Forefront Research Center, University of Osaka, 1-1 Machikaneyama-cho, Toyonaka, Osaka 560-0043, Japan}
\affiliation{Department of Physics, University of Osaka, 1-1 Machikaneyama-cho, Toyonaka, Osaka 560-0043, Japan}
\author{Hidetomo Usui}
\affiliation{Department of Applied Physics, Shimane University, 1060 Nishikawatsu-cho, Matsue, Shimane 690-8504, Japan}
\author{Kazuhiko Kuroki}
\affiliation{Department of Physics, University of Osaka, 1-1 Machikaneyama-cho, Toyonaka, Osaka 560-0043, Japan}

\date{\today}

\begin{abstract}
We theoretically propose a possibility of realizing high $T_c$ superconductivity having $s\pm$-wave symmetry in the heavily hole-doped infinite layer nickelates La$_{1-x}$Sr$_x$NiO$_2$.  We consider situations where the original $P4/mmm$ symmetry of LaNiO$_2$ is maintained even for a significant amount of Sr substitution by growing thin films on substrates having tetragonal symmetry. Considering such cases is indeed justified by our phonon calculations. For electron configurations somewhat close to $d^8$, the interaction between the $d_{x^2-y^2}$ band and the other $3d$ bands that lie just below the Fermi level results in an enhancement of superconductivity where the sign of the gap function is reversed between the former and the latter bands. 
\end{abstract}

\pacs{74.20.Mn,74.70.−b}
\maketitle
\section{Introduction}
The discovery of superconductivity in thin films of lightly hole doped  infinite layer nickelates $RE_{1-x}AE_{x}$NiO$_2$ ($RE=$La, Pr, Nd, $AE=$Ca, Sr)~{\cite{Hwang,Hwang2,Ariando,AriandoCa,OsadaSr,NomuraReview}} has now led to the ``nickel age''\cite{Norman} of unconventional superconductivity. In fact, last year, superconductivity with a maximum $T_c$ exceeding 80 K was discovered in a bilayer Ruddlesden-Popper nickelate La$_3$Ni$_2$O$_7$~\cite{MWang} under pressure, confirmed by the experiments that followed~\cite{Zhang2024,ChengJG,QiYanpeng,ChengJG2,sakakibara4310}.  Subsequently, superconductivity with $T_c= 20\sim 30$ K was discovered in a trilayer version of the nickelate La$_4$Ni$_3$O$_{10}$ also under pressure~\cite{sakakibara4310}, which was further confirmed by other experiments~\cite{2311.05453,La4Ni3O10Nature,PhysRevX.15.021005}. It is worth mentioning that, before entering the nickel age, theoretical studies on the mechanism of unconventional high $T_c$ superconductivity in specific materials have been performed only after their experimental discoveries, but for the bilayer nickelate La$_3$Ni$_2$O$_7$, two of the present authors have discussed the possibility of superconductivity in this material even before its experimental discovery~\cite{Nakata}, and for the trilayer La$_3$Ni$_2$O$_7$, the structural transition to tetragonal symmetry under pressure as well as the occurrence of superconductivity with a $T_c$ comparable to those of the relatively low $T_c$ cuprates was predicted theoretically by the present authors~\cite{sakakibara4310} also before the experimental observations.

As for the infinite layer nickelates, several authors, including us, have theoretically proposed a cuprate-like $d$-wave pairing scenario, since the electron configuration is considered to be close to $d^9$ as in the cuprates~\cite{SakakibaraNi,Thomale,Kitatani,Kitatani2}.
On the other hand, in a previous study, two of the present authors proposed a possibility of realizing $s\pm$-wave superconductivity in the infinite layer nickelates with possibly an even higher $T_c$ than for the $d$-wave pairing in the low doping regime~\cite{Kitamine}. Here, $s\pm$-wave means that the signs of the superconducting gap function between $3d_{x^2-y^2}$ and the other four $3d$ orbitals are reversed. In Ref.~\cite{Kitamine}, the possible existence of residual hydrogen was assumed, as suggested theoretically~\cite{Held,Held2}, so that the electron configuration is closer to $d^8$ than in the absence of the hydrogen. When the energy level offsets $\Delta E$ between $3d_{x^2-y^2}$ orbitals and the other $3d$ (especially $3d_{3z^2-r^2}$) orbitals are large, as in the infinite-layer case, where there are no apical oxygens, the electron configuration somewhat close to $d^8$ would result in a situation where the four $3d$ bands are incipient~\cite{incipient}, 
which leads to a high $T_c$ $s\pm$-wave superconductivity, as suggested previously by two of the present authors~\cite{Kitamine,KitaminePressure}. The basic idea comes from the mathematical equivalence between two-orbital and bilayer Hubbard models~\cite{Shinaoka,Yamazaki}, where the interorbital energy level offset $\Delta E$ in the former is transformed to the interlayer hopping $t_\perp$ in the latter. The experimental circumstances regarding the possibility of residual hydrogen is controversial : there is a study which suggests the residual hydrogen playing an important role~\cite{Liang}, while several studies rule out such a possibility~\cite{PhysRevLett.133.066503,AriandoNoHyd2}.

In the present study, we theoretically seek for another possibility of attaining $s\pm$-wave high $T_c$ superconductivity in the infinite layer nickelates by a realization of electron configuration somewhat close to $d^8$. Namely, we consider heavily doping holes in the original material (say LaNiO$_2$) by largely substituting the rare earth elements (such as La) with alkaline-earth elements (such as Sr or Ba). 
The end materials SrNiO$_2$ and BaNiO$_2$ are known to have orthorhombic crystal structures~\cite{srnio2structure,BaNiO2}, different from the one with tetragonal ($P4/mmm$) symmetry of LaNiO$_2$, but if the $P4/mmm$ symmetry can be maintained even for (La,Sr)NiO$_2$ with significant Sr content by growing thin films on substrates having tetragonal symmetry, there may be a chance of realizing $s\pm$-wave superconductivity with a $T_c$ possibly higher than in the low doping regime. Phonon calculations in the present study indeed suggest such a situation is plausible.

In Ref.~\cite{Kitamine}, a rigid band was assumed in the entire doping regime, and also a fixed set of electron-electron interaction parameters was used. However, since the interorbital energy level offset $\Delta E$ and the interorbital interactions are both crucial in the $s\pm$-wave superconductivity, in the present study we employ the virtual crystal approximation (VCA) to take into account the variance of bands, especially $\Delta E$, upon increasing the substitution content $x$, and also determine the electron-electron interaction values by using constrained random phase approximation (cRPA)~\cite{cRPA} for each $x$ and substrate combination. Our analysis using fluctuation exchange approximation suggests that superconductivity with relatively high $T_c$ is possible, especially when substrates with small lattice constants are used.

\section{Method}
First, we perform first-principles calculation to obtain the band structure of {La$_{1-x}AE{_x}$}NiO$_2$($AE$=Sr, Ba) using the {\footnotesize QUANTUM ESPRESSO} code~\cite{QE}. We assume cases where the in-plane lattice constants are determined by the substrate. We consider three possibilities for the substrate : SrTiO$_3$, (LaAlO$_3$)$_{0.3}$(Sr$_2$TaAlO$_6$)$_{0.35}$ (LSAT), and LaAlO$_3$, whose lattice constants $a$ are 3.905 \AA, 3.868 \AA, and 3.790 \AA, respectively.  We use the crystal structure optimized within this restriction. 
Although the effect of the substrates is complex in reality due to, e.g., charge transfer at the interface, we take into account, for simplicity, only the effect of constraining the in-plane lattice constant to that of the substrate.
Here we briefly comment on the lattice mismatch for the cases considered. For the non-doped bulk LaNiO$_2$, the experimental value of the lattice constant is $a=3.958$ \AA~\cite{Hayward}, and $a$  obtained by fully-relaxing the lattice structure within the first principles calculation is $a=3.937$ \AA, in nice agreement (the error is less than 1~\%).
The lattice mismatches between the experimental value of the bulk lattice constant and the substrates are $1.3, 2.3$ and $4.2$~\% for SrTiO$_3$, LSAT, and LaAlO$_3$ cases, respectively.
Moreover, the fully-relaxed first principles value of $a$ tends to be reduced by Sr doping
(e.g., $a=3.910$ \AA~at $x=0.5$ and $a=3.855$ \AA~at $x=1.0$).
Hence, growing thin films on these substrates can be considered as realistic from the viewpoint of the lattice mismatch, especially  in the heavily doped regime.
As for the comparison between BaNiO$_2$ and SrNiO$_2$, the former has a slightly larger value of the fully-relaxed  lattice constant ($a=3.916$ \AA).
The variance of the lattice constant and the mismatch against the doping ratio are provided in detail in the supplemental material~\cite{SM}. 

We also perform phonon calculation using the density functional perturbation theory.
We take a $3 \times 3 \times 3$ $q$-mesh.  We also adopt VCA to take into account the effect of the partial substitution. 
The Perdew-Burke-Ernzerhof parametrization of the generalized gradient approximation (PBE-GGA)~\cite{PBE-GGA}
and the scalar-relativistic version of the optimized norm-conserving Vanderbilt pseudopotentials~\cite{ONCVP} taken from PseudoDojo~\cite{Dojo} are used.
We take 100 Ry plane-wave cutoff energy, a 12 $\times$ 12 $\times$ 12 $k$-mesh, and an energy width of 0.02 Ry for Gaussian smearing.

We then extract maximally localized Wannier functions~\cite{Marzari,Souza} using the {\footnotesize RESPACK} code~\cite{respack0,respack,respack1,respack2,respack3,respack4,respack5,wan2respack}, 
by which we also obtain the hopping parameters among the Wannier functions. 
We have constructed a five-orbital model consisting of all the Ni $3d$ orbitals
and performed cRPA calculation to determine the Coulomb and exchange interaction parameter of the model,
where the doping effects on these parameters are small.
In the actual many-body calculation, we consider only the on-site intra- and interorbital repulsions, Hund's coupling, and the pair hopping terms. 
Important interaction parameter values are given in Table \ref{tab1}, where we find  that they do not strongly depend on the doping ratio or the substrate.

\begin{table}[!h]
\caption{Interaction parameters of onsite Coulomb repulsion $U, U'$,
Hunds' coupling $J$ and pair-hopping $J'$ evaluated with cRPA. The orbital indices $l,m=1,2,3,4,5$ indicate
the $d_{x^2-y^2}, d_{3z^2-r^2}, d_{xy}, d_{yz}, d_{zx}$ orbitals, respectively. For simplicity, we abbreviate four-orbital-index-representation
of partially screened interaction integral $V$, namely, $V_{llmm} \rightarrow U'_{lm}$ and 
$V_{lmlm}(=V_{lmml}) \rightarrow J_{lm}$. Note that $U'_{lm}=U'_{ml}, J_{lm}=J_{ml}$.
The units of all parameters are eV.  }
\label{tab1}
\begin{tabular}{c|cc|cc|cc}
\hline
\hline
&  $x=0.5$& (STO)&$x=1.0$ & (STO) &$x$=0.5 & (LSAT) \\
\hline
 $l,m$ & $U,U'$ & $J, J'$ & $U,U'$ & $J, J'$& $U,U'$ & $J, J'$ \\
\hline
 1,1 & 3.26 & --      & 3.34& -- & 3.23 & --\\
 1,2 & 1.88 & 0.64  & 2.42& 0.64 & 1.86& 0.64\\
 1,3 & 2.60 & 0.38  & 2.66 & 0.32 & 2.56&0.38\\
 1,4 & 2.15 & 0.65  & 2.52 & 0.58  & 2.12& 0.65\\
 1,5 & 2.15 & 0.65  & 2.52 & 0.58 & 2.12& 0.65\\
 2,2 & 3.50 & --      & 4.64& -- & 3.52&--\\
 2,3 & 1.97 & 0.63  & 2.44 & 0.61& 1.94& 0.64\\
 2,4 & 2.51 & 0.53  & 3.29 & 0.53 & 2.49&0.54\\
 2,5 & 2.51 & 0.53  & 3.29 & 0.53 & 2.49&0.54\\
 3,3 & 3.58 & --     & 3.35 & --&  3.52 &--\\
 3,4 & 2.27 & 0.69 & 2.55 & 0.58 &2.23 &0.69\\
 3,5 & 2.27 & 0.69  & 2.55 & 0.58 &2.23 &0.69\\
 4,4 & 3.69 & --  & 4.19 & -- & 3.66 &--\\
 4,5 & 2.25 & 0.68  & 2.79 & 0.63 & 2.22& 0.68\\
 5,5 & 3.69 & --  &4.19 & -- & 3.66 &--\\
 \hline
\end{tabular}
\end{table}

We explore the possibility of superconductivity 
for the obtained low-energy five-orbital model within the fluctuation-exchange (FLEX) approximation~\cite{Bickers,Bickers1991}. 
We calculate the self-energy induced by the spin-fluctuation formulated as shown in the literatures~\cite{Lichtenstein,mFLEX1,mFLEX2} in a self-consistent calculation.
The explicit formulae of the irreducible, spin, and charge susceptibilities 
describing the fluctuations are shown in Eqs.(2)-(4) of Ref.~\cite{Sakakibara2}.
The real part of the self-energy at the lowest Matsubara frequency is subtracted in the same manner with Ref.~\cite{Ikeda_omega0}
to maintain the band structure around the Fermi level obtained by first-principles calculation.
In the present cases where the level offset between the $d_{x^2-y^2}$ and other four orbitals is large, large $x$ region corresponds to the case where the $d_{x^2-y^2}$ band is nearly empty and the other bands are nearly fully filled. This case is closer to a band insulator rather than a Mott insulator. We believe that in such cases, the electron correlation would not be so strong as to make FLEX unreliable.

The obtained Green's function and the pairing interaction, mediated mainly by spin fluctuations, are plugged into the linearized Eliashberg equation.
Since the eigenvalue $\lambda$ of the linearized Eliashberg equation reaches unity at $T=T_c$, 
we adopt it as a measure of superconductivity at a fixed temperature, $T=0.01$ eV. 
For convenience, we will call the eigenfunction (with the largest eigenvalue) of the linearized Eliashberg equation at the lowest Matsubara frequency $i\omega$(=$i\pi k_{\rm B}T$) the ``superconducting gap function''. We take a 16$\times$16$\times$4 $k$-point mesh and 4096 Matsubara frequencies for the FLEX calculation.

\begin{figure}
	\includegraphics[width=9cm]{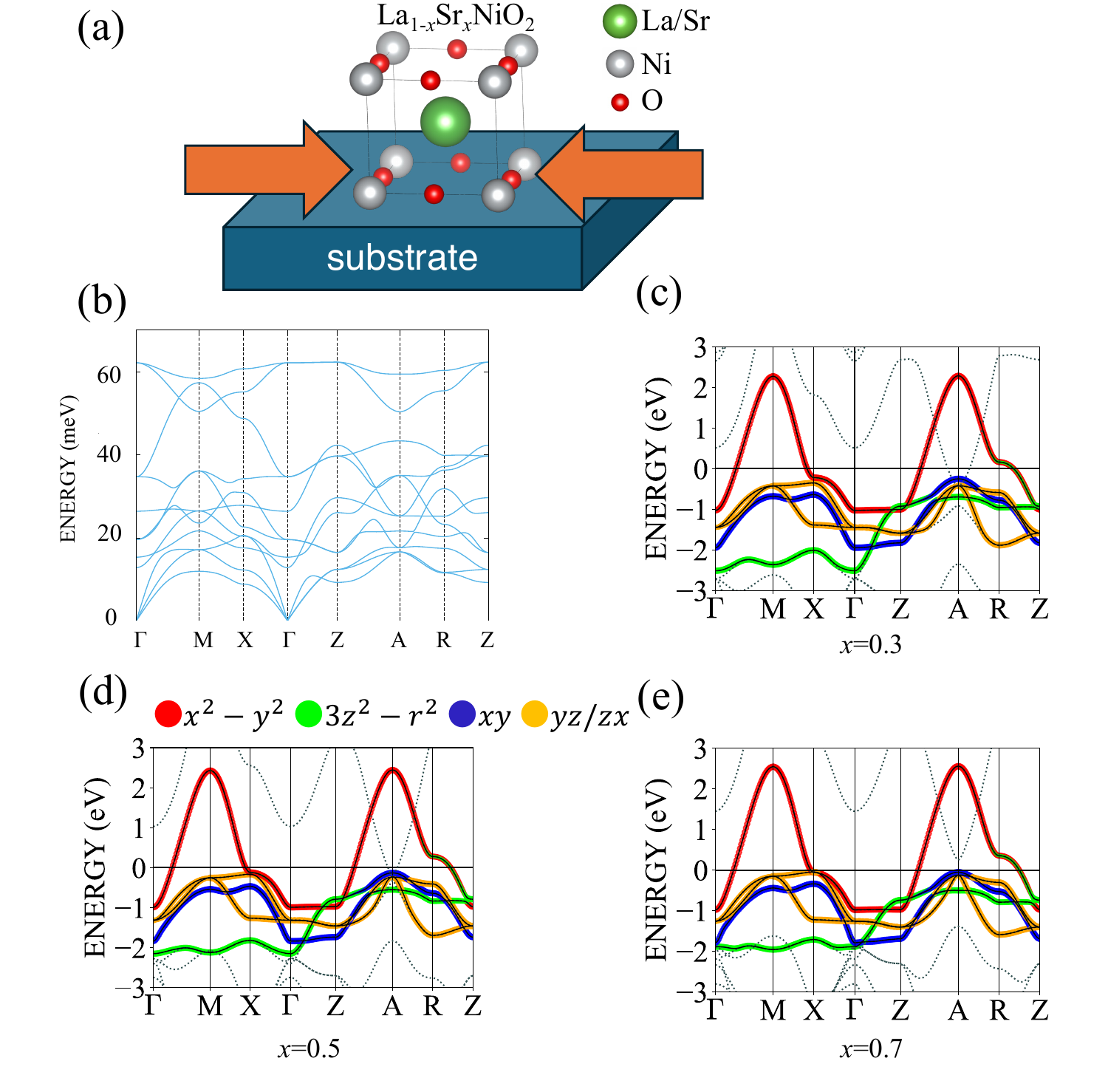}	
	\caption{(a) An image of the system considered in the present study. (b) The phonon dispersion of La$_{1-x}$Sr$_x$NiO$_2$ for $x=0.5$, fixing the in-plane lattice constant to that of the SrTiO$_3$ substrate. The corresponding electronic band dispersion for (c) $x=0.3$, (d) $x=0.5$, and (e) $x=0.7$. The strength of the Wannier orbital characters are shown in (c)-(d) with the thickness of the color coded line.
	}
	\label{fig1}
\end{figure}

\begin{figure}
	\includegraphics[width=9cm]{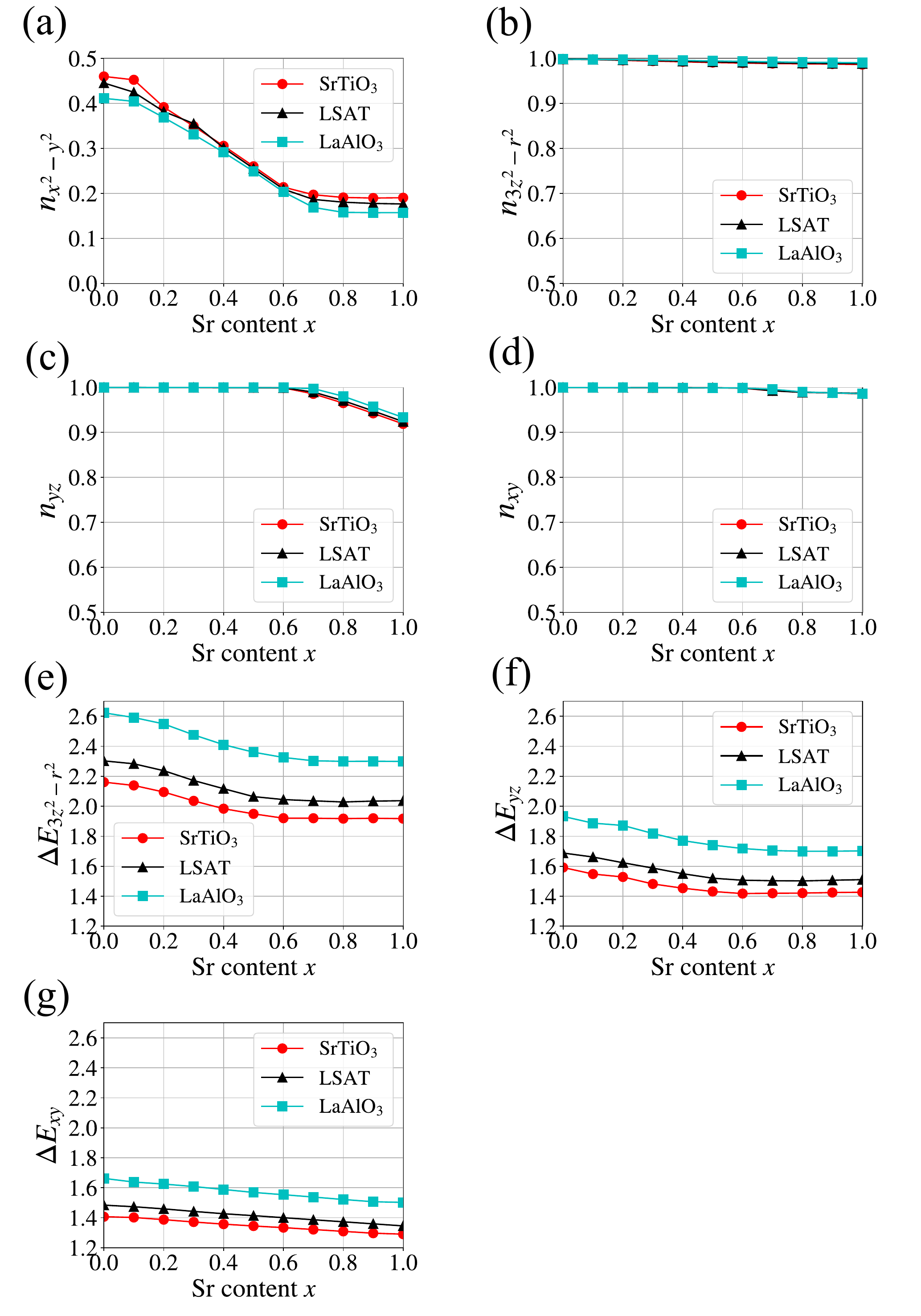}	
	\caption{(a)--(d) the number of electrons per site per spin ($n=1$ means that the orbital is fully filled)  for each orbital against the Sr content $x$ for the three substrates. (a) $d_{x^2-y^2}$ (b) $d_{3z^2-r^2}$ (c) $d_{yz}$ (d) $d_{xy}$ orbitals. (e)--(g) the level offset between $d_{x^2-y^2}$ and the other orbitals against the Sr content $x$ for the three substrates. (e) $d_{3z^2-r^2}$ (f) $d_{yz}$ (g) $d_{xy}$ orbitals. 
	}
	\label{fig2}
\end{figure}

\begin{figure}
	\includegraphics[width=9cm]{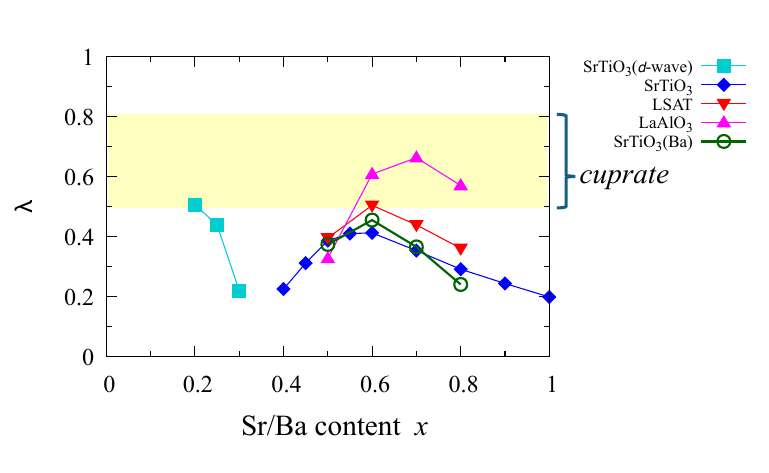}	
	\caption{The eigenvalue of the Eliashberg equation plotted against the Sr/Ba content $x$ for the three substrates. The pairing symmetry having the largest $\lambda$ is $d$-wave in the small $x$ regime (calculated only for SrTiO$_3$ substrate), while it is $s\pm$-wave for $x>0.3$.
	The yellow hatched region indicates the range of $\lambda$ calculated for the cuprates~\cite{Sakakibara1,mrpa2} by FLEX at $T=0.01$ eV.
	}
	\label{fig3}
\end{figure}

\begin{figure}
	\includegraphics[width=9cm]{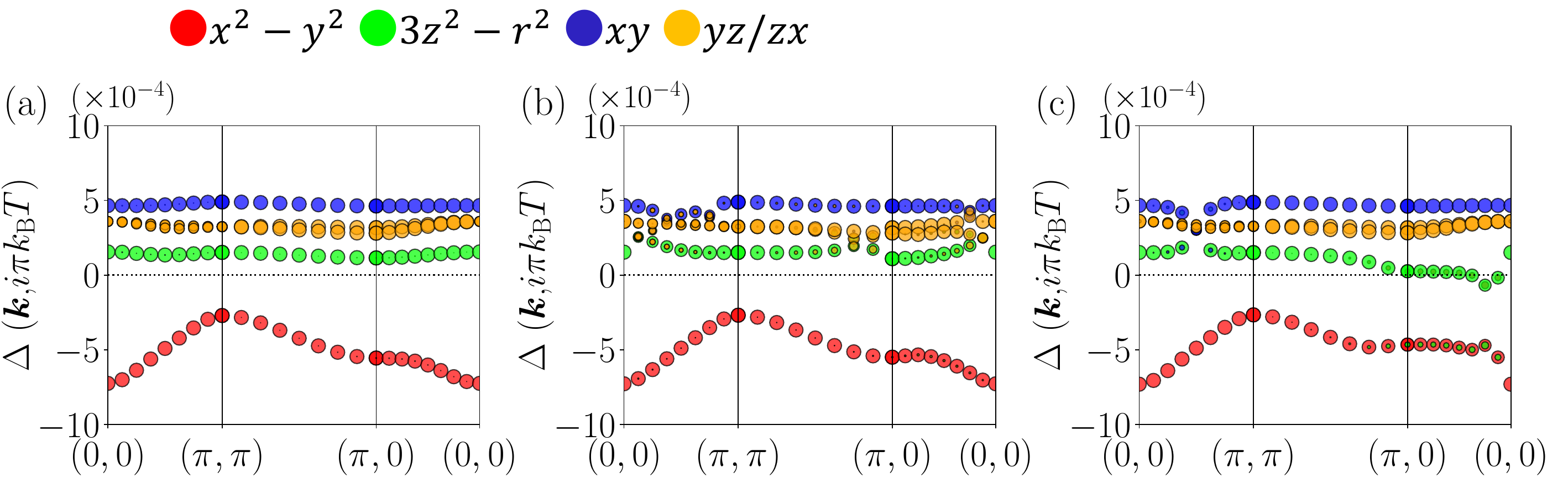}	
	\caption{The gap function (eigenfunction of the Eliashberg equation) in the orbital representation for the $x=0.6$ case of La$_{1-x}$Sr$_x$NiO$_2$ with the LSAT substrate, plotted against the in-plane wave vector $(k_x,k_y)$ at the out-of-plane wave vectors (a) $k_z=0$, (b) $k_z=\pi/2$, and (c) $k_z=\pi$. }
	\label{fig4}
\end{figure}

\section{Results and Discussions}
As for the stability of the lattice structure of La$_{1-x}$Sr$_x$NiO$_2$, our phonon calculations show that, interestingly, the $P4/mmm$ symmetry of the original LaNiO$_2$ is dynamically stable for the entire Sr content range of $0\leq x \leq 1$ for all the three substrates considered here. 
In Fig.~\ref{fig1} (b), we present the phonon dispersion for the case of $x=0.5$ assuming a SrTiO$_3$ substrate, where imaginary phonon frequencies do not appear. The rest of the phonon calculations are presented in the supplemental material~\cite{SM}. 

Our phonon calculations justify adopting the $P4/mmm$ symmetry of the lattice structure for the electronic band structure calculation and model construction for all the cases considered in the present study. In Figs.~\ref{fig1}(c)--1(e), we show the band structure of La$_{1-x}$Sr$_x$NiO$_2$ adopting the in-plane lattice constant of the SrTiO$_3$ substrate for  various Sr content $x$. While the Fermi level intersects the $d_{x^2-y^2}$ band, it stays somewhat away from the  $d_{3z^2-r^2}$ band for all $x$ because the apical oxygen is absent. On the other hand, the Fermi level approaches the top of the $t_{2g}$ bands as $x$ is increased, and eventually intersects those bands for too large $x$. 
The band structures for the other substrates are presented in the supplemental material~\cite{SM}.

In Fig.~\ref{fig2}, we plot the number of electrons and $\Delta E$ (the energy level offset from the $d_{x^2-y^2}$ orbital) of each orbital against the Sr content $x$ for various substrates. It can be seen that $\Delta E$ depends on the substrate with a larger $\Delta E$ for a shorter lattice constant. This is because when the in-plane lattice constant is shorter, the energy level of the $d_{x^2-y^2}$ orbital, elongated toward the in-plane oxygens, is pushed up. 
Also, the relatively large variation of $\Delta E$ against $x$ indicates that the rigid band picture does not hold.

As for the number of electrons in each orbital, obviously, that of the $d_{x^2-y^2}$ orbital decreases as $x$ is increased, but a more important point is that at certain point that of the $d_{yz}$ (and equivalently $d_{xz}$) orbital starts to decrease, indicating that the Fermi level has started intersecting these bands. This critical point of $x$ is larger for substrates with shorter lattice constant, reflecting the larger values of $\Delta E$ between the $d_{x^2-y^2}$ and $d_{xz/yz}$ orbitals.

In Fig.~\ref{fig3}, we plot the eigenvalue of the linearized Eliashberg equation against the Sr content $x$  for the three substrates. For comparison, we also show the range of the eigenvalue calculated for various cuprates within the same formalism. For $x\geq 0.4$, the dominant pairing symmetry becomes $s\pm$-wave, and the eigenvalue is maximized around $x\simeq 0.5$ -- $0.7$, depending on the substrate. 
As mentioned before, by $s\pm$-wave, we mean that the sign of the gap function is reversed between the $d_{x^2-y^2}$ and the other orbitals, as is depicted in Fig.~\ref{fig4} for the case of $x=0.6$ and the LSAT substrate. Only for the SrTiO$_3$ substrate, we calculate the eigenvalue at a low Sr content of $x=0.2$, for which the dominant pairing symmetry is $d$-wave. We note here that for small $x$ (where $d$-wave dominates), the rare earth $5d$ bands are known to intersect the Fermi level, and in such a case, a more accurate evaluation of the electron-electron interaction by cRPA requires including those bands~\cite{SakakibaraNi}. Since this is not done in the present study, the interaction values are underestimated in the low doping regime, which results in an overestimation of $\lambda$. 
The difference in $\lambda$ between the cases of $AE={\rm Sr}$ and Ba is small, which may be due to the fact that the in-plane lattice constant is fixed at the value of the substrate. 
We have also calculated Stoner factor $U\chi(\bm{q},i\omega)$ to evaluate the magnetic instability. In the STO (LSAT) case, the values are 0.86, 0.87, 0.90 (0.85, 0.87, 0.89) for the substitution ratios $x=0.5, 0.6, 0.7$, respectively.  These values are smaller than the corresponding values for actual high $T_c$ superconductors such as the cuprates (around 0.97--0.99) and the bilayer nickelates (around 0.94--0.96), suggesting that the tendency towards magnetism is weak compared to these existing superconductors.

For substrates with shorter lattice constants, $\lambda$ is maximized at a larger $x$ with a larger maximum value. This is because for shorter lattice constants, $\Delta E$ is larger  (Fig.~\ref{fig2}) so that the Fermi level starts to intersect the $d_{xz/yz}$ bands at a larger $x$, for which the superconductivity is degraded due to strong renormalization effects, and also the large $\Delta E$ itself favors high $T_c$~\cite{Kitamine,KitaminePressure,Yamazaki}.

We have subtracted the real part of the self-energy to prevent double counting of the effect of the electron-electron interaction, but there can be other ways to cope with this problem. In particular, the energy level offset between different orbitals may depend on the adopted approach. In order to check how the difference in the energy level offset would affect superconductivity, we have adopted GGA$+U$ to perform DFT calculation and hence the model construction. 
We use the rotationally invariant formulation introduced by Dudarev {\it et al.}~\cite{ldau}, adopting the ortho-atomic type of Hubbard projections.
The result is summarized in the supplemental material~\cite{SM}. 
Compared to the GGA case, the level offset between the $d_{x^2-y^2}$ and the other orbitals is enlarged. This effect makes the lower $d$-bands lie farther away from the Fermi level.
As a result, the region in which the lower $d$-bands become incipient shifts towards heavier doped region. This difference can change optimally hole-doped region for $s\pm$-wave superconductivity but the theoretical expectation that the heavily hole doped infinite layered nickelate can exhibit superconductivity remains, although there is an ambiguity in the amount of holes required.

\section{Summary}
We have theoretically examined the possibility of realizing high $T_c$ superconductivity having $s\pm$-wave symmetry in the heavily hole-doped infinite layer nickelates La$_{1-x}$Sr$_x$NiO$_2$. We have considered situations where the original $P4/mmm$ symmetry of LaNiO$_2$ is maintained for  significant amount of Sr substitution by growing thin films on substrates having tetragonal symmetry. We have shown that considering such cases is justified by performing phonon calculations for the crystal structures obtained by fixing the in-plane lattice constants to those of several substrates.

For electron configurations somewhat close to $d^8$, the interaction between the $d_{x^2-y^2}$ band and the other $3d$ incipient bands that lie just below the Fermi level results in an enhancement of superconductivity where the sign of the gap function is reversed between the $d_{x^2-y^2}$ and the incipient bands. The strong enhancement of superconductivity can be attributed to the large energy level offset between the $d_{x^2-y^2}$ and the other orbitals due to the absence of the apical oxygens, as has been pointed out in previous studies~\cite{Kitamine,KitaminePressure,Yamazaki}. To investigate this effect in a quantitative manner, we have employed VCA to take into account the variance of bands upon increasing the substitution content $x$, and have also determined the electron-electron interaction values by cRPA for each $x$ and substrate combination. Our FLEX analysis has shown that superconductivity with relatively high $T_c$ is possible, especially when substrates with small lattice constants are adopted.

\begin{acknowledgments}
We are supported by JSPS KAKENHI Grants No. JP22K03512, No. JP25K08459 (H. S.), No. JP22K04907 (K. K.), and No. JP24K01333. The computing resource is supported by 
the supercomputer system (system-B) in the Institute for Solid State Physics, the University of Tokyo, 
and the supercomputer of Academic Center for Computing and Media Studies (ACCMS), Kyoto University.
\end{acknowledgments}

\bibliography{srnio2}

\begin{thebibliography}{60}
\bibitem [{inc()}]{incipient}%
  \BibitemOpen
  \href@noop {} {}\bibinfo {note} {The incipient band in the narrow sense of
  the term means that the band does not intersect the Fermi level. In this
  paper, we use the word ``incipient band'' in a more loose sense. Namely, we
  call the band incipient when the top (or bottom) of the band is located near
  the Fermi level, including the case when the band intersects the Fermi
  level.}\BibitemShut {Stop}%
\end{thebibliography}

\end{document}